\RequirePackage{lineno}

\documentclass[twocolumn,showpacs,aps,prd,superscriptaddress]{revtex4}

\usepackage{graphicx}
\usepackage{dcolumn}
\usepackage{amsmath}
\usepackage{epsfig}

\usepackage{enumitem}

\input babarsym

\begin{document}
\begin{flushleft}
\babar-PUB-14/012\\
SLAC-PUB-16230\\[10mm]
\end{flushleft}
\title{
{\large \bf
Search for a light Higgs resonance in radiative decays of the $\mathbf{\Upsilon(1S)}$ with a charm tag
} 
}

%% author list as of 04-Nov-2014 (293 authors)
%
\author{J.~P.~Lees}
\author{V.~Poireau}
\author{V.~Tisserand}
\affiliation{Laboratoire d'Annecy-le-Vieux de Physique des Particules (LAPP), Universit\'e de Savoie, CNRS/IN2P3,  F-74941 Annecy-Le-Vieux, France}
\author{E.~Grauges}
\affiliation{Universitat de Barcelona, Facultat de Fisica, Departament ECM, E-08028 Barcelona, Spain }
\author{A.~Palano$^{ab}$ }
\affiliation{INFN Sezione di Bari$^{a}$; Dipartimento di Fisica, Universit\`a di Bari$^{b}$, I-70126 Bari, Italy }
\author{G.~Eigen}
\author{B.~Stugu}
\affiliation{University of Bergen, Institute of Physics, N-5007 Bergen, Norway }
\author{D.~N.~Brown}
\author{L.~T.~Kerth}
\author{Yu.~G.~Kolomensky}
\author{M.~J.~Lee}
\author{G.~Lynch}
\affiliation{Lawrence Berkeley National Laboratory and University of California, Berkeley, California 94720, USA }
% ...
\collaboration{The \babar\ Collaboration}
\noaffiliation

\begin{abstract}
A search is presented for the decay $\Upsilon(1S)\to\gamma A^0$, $A^0\to c\bar{c}$, where $A^0$ is a candidate for the \ensuremath{C\!P}\xspace-odd Higgs boson of the next-to-minimal supersymmetric standard model. The search is based on data collected with the $\mbox{\slshape B\kern-0.1em{\smaller A}\kern-0.1em
    B\kern-0.1em{\smaller A\kern-0.2em R}}$\ detector at the $\Upsilon(2S)$ resonance. A sample of $\Upsilon(1S)$ mesons is selected via the decay $\Upsilon(2S)\to\pi^+\pi^-\Upsilon(1S)$. The $A^0\to c\bar{c}$ decay is identified through the reconstruction of hadronic $D^0$, $D^+$, and $D^*(2010)^+$ meson decays. No significant signal is observed. The measured 90\% confidence-level upper limits on the product branching fraction $\mathcal{B}(\Upsilon(1S)\to \gamma A^0) \times \mathcal{B}(A^0 \to c\bar{c}$) range from $7.4 \times 10^{-5}$ to $2.4 \times 10^{-3}$ for $A^0$ masses from 4.00 to 8.95 GeV/$c^2$ and 9.10 to 9.25 GeV/$c^2$, where the region between 8.95 and 9.10 GeV/$c^2$ is excluded because of background from $\Upsilon(2S)\to\gamma \chi_{bJ}(1P)$, $\chi_{bJ}(1P)\to \gamma \Upsilon(1S)$ decays.
\end{abstract}

\pacs{12.15.Ji, 12.60.Fr, 13.20.Gd, 14.80.Da}

\maketitle

The next-to-minimal supersymmetric standard model (NMSSM) is an appealing extension of the standard model (SM). It solves the $\mu$-problem of the minimal supersymmetric standard model and the hierarchy problem of the SM~\cite{ref:radups, ref:NMSSMreview}. The NMSSM has a rich Higgs sector of two charged, three neutral $\CP$-even, and two neutral $\CP$-odd bosons. Although the Higgs boson discovered at the CERN LHC~\cite{ref:ATLAShiggs1, ref:CMShiggs1} is consistent with the SM Higgs boson, it can also be interpreted as one of the heavier Higgs bosons of the NMSSM~\cite{ref:NMSSMLHC}. The least heavy of the NMSSM Higgs bosons, denoted $A^0$, could be light enough to be produced in the decay of an $\Upsilon$ meson~\cite{ref:radups, ref:wilczek}.

In the context of type I or type II two-Higgs-doublet models, the branching fractions of the $A^0$ depend on the $A^0$ mass and the NMSSM parameter $\tan\beta$~\cite{ref:update}. Below the charm mass threshold, the $A^0$ preferentially decays into two gluons if $\tan\beta$ is of order 1, and to \ssbar or to \mumu if $\tan\beta$ is of order 10. Above the charm mass threshold, the $A^0$ decays mainly to \ccbar for $\tan\beta$ of order 1 and to \tautau for $\tan\beta$ of order 10. \babar\ has already ruled out much of the NMSSM parameter space for $A^0$ masses below the charm mass threshold through searches for $A^0\to\mumu$~\cite{ref:mumu,ref:mumu2} and for $A^0\to gg$ or \ssbar~\cite{ref:ggss}. Above the charm mass threshold, \babar\ has ruled out some of the parameter space for high $\tan\beta$ with the $A^0\to\tautau$ searches~\cite{ref:tautau,ref:tautau2}. None of the searches from \babar\ have observed a significant signal, nor have the searches in leptonic channels from the CMS and CLEO~\cite{ref:CMSmumu,ref:CMSmumu2,ref:CLEOa1} Collaborations. The $A^0\to\ccbar$ channel is one of the last channels that has not yet been explored.

We report a search for the decay $\OneS\to\gamma A^0$, $A^0\to\ccbar$ with $A^0$ masses ranging between 4.00 and 9.25~\gevcc. An \OneS decay is tagged by the presence of a pion pair from $\TwoS\to \pipi \OneS$. An $A^0\to\ccbar$ decay is tagged by the presence of at least one charmed meson such as a \Dz, a \Dp, or a $D^*(2010)^+$. Therefore candidates are constructed from the combination of a photon, a $D$ meson, and a dipion candidate. An exclusive reconstruction of the $A^0$ is not
attempted. Instead, a search is performed in the spectrum of the invariant mass of the system that recoils against the dipion-photon system. The analysis is therefore sensitive to the production of any charm resonance produced in the radiative decays of the \OneS meson. 

The data were recorded with the \babar\ detector at the \pep2\ asymmetric-energy \epem\ collider at the SLAC National Accelerator Laboratory. The \babar\ detector is described in detail elsewhere~\cite{ref:detector,ref:detector2}. We use 13.6\invfb of ``on-resonance" data collected at the \TwoS resonance, corresponding to $(98.3\pm0.9) \times 10^6$ \TwoS mesons \cite{Lees:2013rw}, which includes an estimated $(17.5\pm0.3) \times 10^6$ $\TwoS\to\pipi\OneS$ decays~\cite{ref:pdg}. The non-\TwoS backgrounds are studied using 1.4\invfb of ``off-resonance'' data collected 30~MeV below the \TwoS resonance.

The EvtGen event generator~\cite{ref:evtgen} is used to simulate the signal event decay chain, $\epem\to\TwoS\to\pipi\OneS, \OneS\to\gamma A^0$, $A^0\to\ccbar$, for $A^0$ masses between 4.0 and 9.0~\gevcc in 0.5~\gevcc steps and for $A^0$ masses of 9.2, 9.3, and 9.4~\gevcc. The $A^0$ decay width is assumed to be 1~\mev. The hadronization of the \ccbar system is simulated using the Jetset~\cite{ref:jetset} program. The detector response is simulated with the GEANT4~\cite{ref:geant4} suite of programs.

Photon candidates are required to have an energy greater than 30~MeV and a Zernike moment $A_{42}$~\cite{ref:Zernike} less than 0.1. The $A_{42}$ selection reduces contributions from hadronic showers identified as photons. Events are required to contain at least one photon candidate. Each photon candidate is taken in turn to represent the radiative photon in the $\OneS\to\gamma A^0$ decays.  We do not select a best signal candidate, neither for the radiative photon nor for the $D$ meson and dipion candidates discussed below, but rather allow multiple candidates in an event.

Events must contain at least one $D$ meson candidate, which is reconstructed in five channels: $D^0 \to K^- \pip$, $D^+ \to K^- \pip \pip$, $D^0 \to K^- \pip \pip \pim$, $D^0 \to \KS \pipi$, and $D^*(2010)^+ \to \pip D^0$ with $D^0 \to K^- \pip \piz$. The $D^0 \to K^- \pip \piz$ decays are reconstructed in the $D^*(2010)^+$ production channel to reduce a large background that would otherwise be present. The inclusion of charge conjugate processes is implied. The \piz candidates are reconstructed from two photon candidates by requiring the invariant mass of the reconstructed \piz to lie between 100 and 160~\mevcc. The \piz candidates do not make use of the radiative photon candidate. The \KS candidates are reconstructed from two oppositely charged pion candidates. Each \KS candidate must have a reconstructed mass within 25\mevcc of the nominal \KS mass~\cite{ref:pdg} and satisfy $d/\sigma_d > 3$, where $d$ is the distance between the reconstructed \epem collision point and the \KS vertex, with $\sigma_d$ the uncertainty of $d$.

The \Dz and \Dp candidates are required to have masses within 20 \mevcc of their nominal masses~\cite{ref:pdg}, corresponding to 3 to 4 standard deviations ($\sigma$) in their mass resolution. When reconstructing $D^*(2010)^+$ candidates, we constrain the $D^0 \to K^- \pip \piz$ candidate mass to its nominal value~\cite{ref:pdg}. The $D^*(2010)^+$ candidate mass distribution has longer tails. The $D^*(2010)^+$ candidates are  required to lie within 5~\mevcc of its nominal mass~\cite{ref:pdg}, corresponding to 10~$\sigma$ in the mass resolution.

\begin{figure}[h]
\includegraphics[width=\columnwidth]{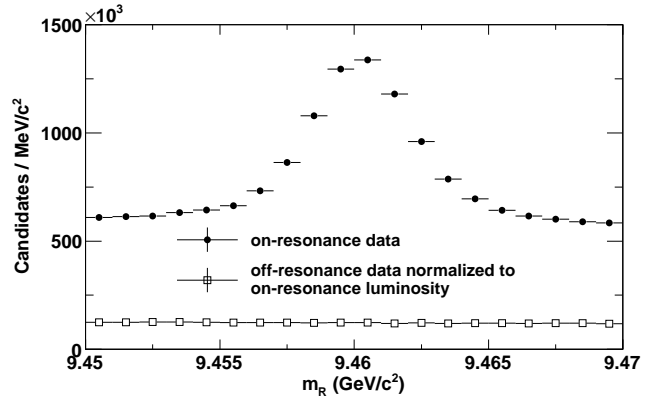}
\caption{
The $m_{R}$ distribution of events with a  dipion, charm, and photon tag before application  of selection criteria based on the BDT output (see text). The solid circles indicate the on-resonance data. The open squares indicate the off-resonance data normalized to the on-resonance luminosity.
}
\label{fig:dpmass}
\end{figure}

Events are required to have at least one dipion candidate, constructed from two oppositely charged tracks. The invariant mass, $m_{R}$, of the system recoiling against the dipion in the $\TwoS\to\pipi\OneS$ transition is calculated by
\begin{equation}
m_{R}^2 = M_{\TwoS}^2 + m_{\pi\pi}^2 - 2M_{\TwoS}E_{\pi\pi},
\end{equation}

\noindent where $m_{\pi\pi}$ is the measured dipion mass, $M_{\TwoS}$ is the nominal \TwoS mass~\cite{ref:pdg}, and $E_{\pi\pi}$ is the dipion energy in the \epem center-of-mass (CM) frame. The two pions in the dipion system are required to arise from a common vertex. Signal candidates must satisfy $9.45 < m_R < 9.47$~\gevcc. Figure \ref{fig:dpmass} presents the distribution of $m_R$ after application of these criteria. A clear peak is seen at the \OneS mass.

All charged tracks and calorimeter clusters other than those used to define the radiative photon, the $D$ meson candidate, and the dipion candidate are referred to as the ``rest of the event''.

The mass of the $A^0$ candidate, $m_X$, is determined from the mass of the system recoiling against the dipion and photon through
\begin{equation}
m_X^2 = (P_{\epem} - P_{\pipi} - P_{\gamma})^2,
\end{equation}

\noindent where $P$ denotes four-momentum measured in the CM frame. The four-momentum of the \epem system is given by $P_{\epem} = (M_{\TwoS},0,0,0)$.

Backgrounds are evaluated using simulated \TwoS and $\epem\to\qqbar$ events, where $q$ is a $u, d, s,$ or $c$ quark. Events with low-energy photons contribute a large background for $m_X$ greater than 7.50~\gevcc. Therefore, the analysis is divided into a low $A^0$ mass region (4.00 to 8.00 \gevcc) and a high $A^0$ mass region (7.50 to 9.25 \gevcc). The definitions of the regions, which overlap, are motivated by the need to have sufficient statistical precision for the background determination in each region.

We train ten boosted decision tree (BDT) classifiers \cite{ref:TMVA} to separate background from signal candidates (two mass regions $\times$ five $D$ channels). The BDTs are trained using samples of simulated signal events, simulated generic \TwoS events, and the off-resonance data. The BDT inputs consist of 24 variables:

\begin{enumerate}[leftmargin=1.4cm]
  \item[1-2.] Event variables:
  	\begin{itemize}[leftmargin=0cm]
		\item number of charged tracks in the event,
		\item number of calorimeter clusters in the event.
  	\end{itemize}
  \item[3-12.] Kinematic variables:
 	\begin{itemize}[leftmargin=0cm]
		\item $m_{R}$,
		\item dipion likelihood (defined later),
		\item $D$ candidate mass,
		\item $D$ candidate momentum,
		\item photon \piz score (defined later),
		\item energy of the most energetic charged track in the rest of the event, calculated using a charged pion mass hypothesis,
		\item energy of the most energetic calorimeter cluster in the rest of the event,
		\item invariant mass of the rest of the event,
		\item CM frame momentum of the rest of the event,
		\item CM frame energy of the rest of the event.
	\end{itemize}
  \item[13-15.] Vertex variables:
  	\begin{itemize}[leftmargin=0cm]
		\item transverse coordinate of a vertex formed using all charged tracks,
		\item longitudinal coordinate of a vertex formed using all charged tracks, 
		\item the $\chi^2$ probability of a vertex fit using all charged tracks.
  	\end{itemize}
  \item[16-18.] Event shape variables:
  	\begin{itemize}[leftmargin=0cm]
		\item the ratio of the second to zeroth Fox-Wolfram moment~\cite{ref:r2all}, calculated using all charged tracks and calorimeter clusters,
		\item sphericity~\cite{ref:spher} of the event,
		\item magnitude of the thrust~\cite{ref:thrust}.
  	\end{itemize}
  \item[19-24.] Opening angles in the CM frame between the
  	\begin{itemize}[leftmargin=0cm]
		\item dipion and photon candidate,
		\item dipion and $D$ candidate,
		\item dipion and thrust axis,
		\item photon and $D$ candidate,
		\item photon and thrust axis,
		\item $D$ candidate and thrust axis.
  	\end{itemize}
\end{enumerate}

\noindent The kinematic variables provide the most separation power for all ten BDTs. The separation power of the other variables depends on the mass region and channel. The vertex variables suppress background without a $D$ meson in the event. The event shape variables suppress $\epem\to\qqbar$ backgrounds.

The dipion likelihood \cite{ref:TMVA} is defined using the opening angle between the two charged pions in the CM frame, the transverse momentum of the pair, the invariant mass of the pair, the larger of the two momenta of the pair, and the $\chi^2$ probability of the pair's vertex fit.

To reject photon candidates from $\piz \to \gamma \gamma$ decays, a likelihood \cite{ref:TMVA} is defined using the invariant mass of the radiative photon candidate and a second photon (if present), and the second photon's CM energy. The lower the likelihood, the more \piz-like the photon pair. The photon \piz score is the minimum likelihood formed from the radiative photon and any other photon in the event excluding photon candidates used to reconstruct the \piz candidate in the $D^0 \to K^- \pip \piz$ decay. 

For each channel and mass range, each BDT output variable is required to exceed a value determined by maximizing the quantity $S/(1.5 + \sqrt{B})$~\cite{ref:Punzi}, where $S$ and $B$ are the expected numbers of signal and background events, respectively, based on simulation.

In the case of events with multiple signal candidates that satisfy the selection criteria, there may be multiple values of $m_X$. Signal candidates that have the same dipion and radiative photon candidate have the same value of $m_X$, irrespective of which $D$ candidate is used. We reject a signal candidate if its value of $m_X$ has already been used.

\begin{figure}[h]
\includegraphics[width=\columnwidth]{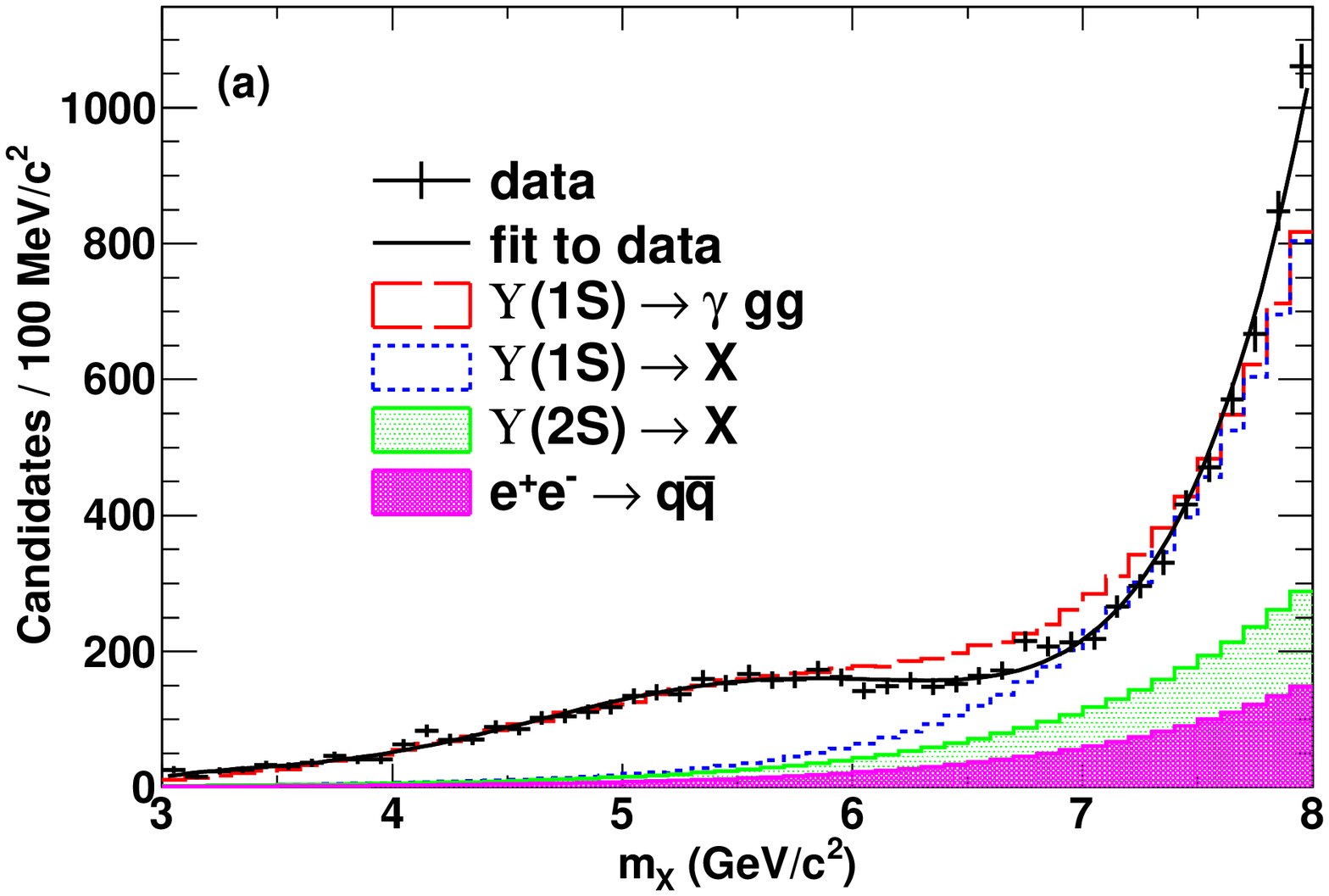}
\includegraphics[width=\columnwidth]{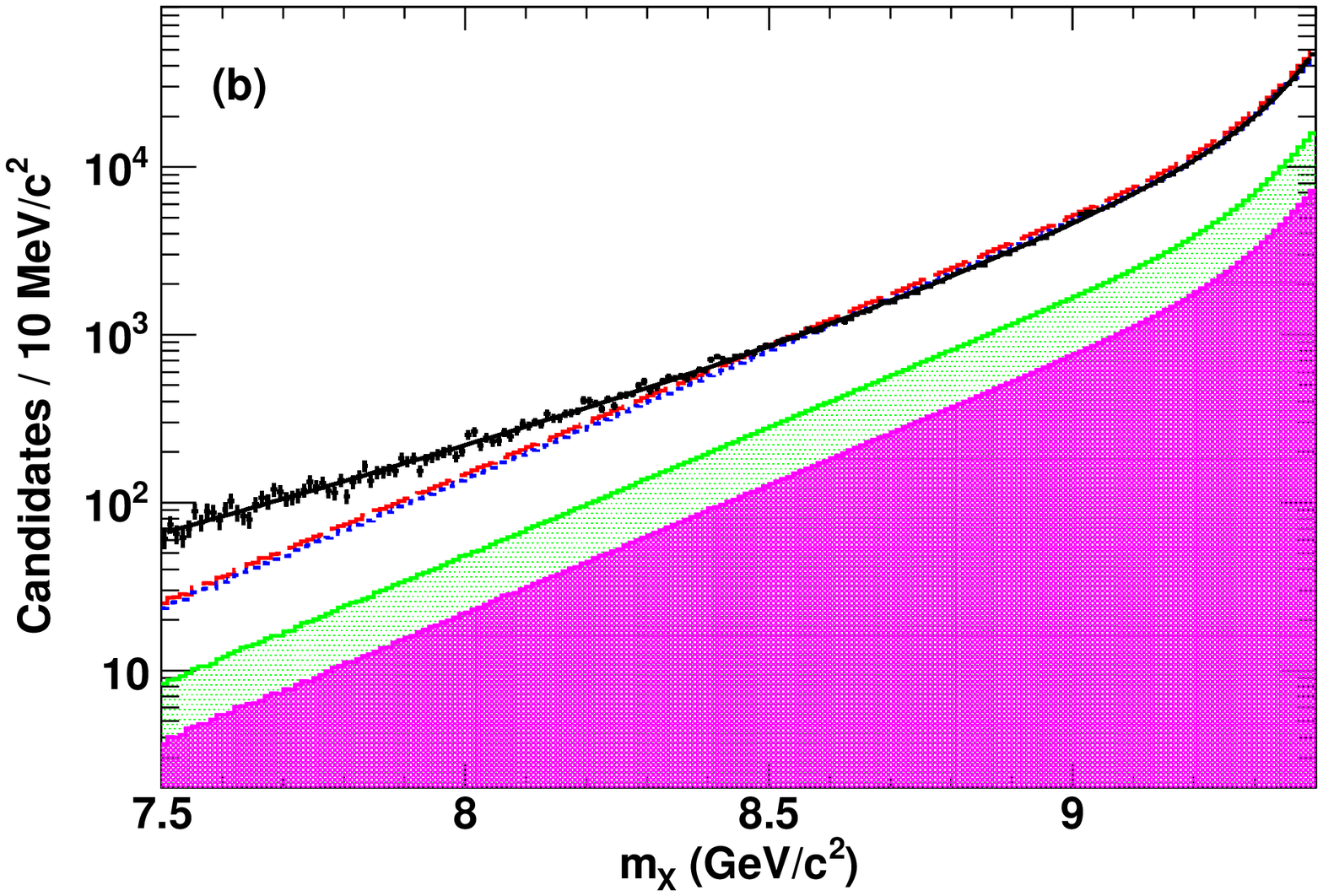}
\caption{
The $m_X$ distributions of signal candidates in the low- (a) and high- (b) mass regions after applying all selection criteria. The points indicate the data.  The solid curve shows the result of a fit to the data under a background-only hypothesis. The colored histograms show the cumulative background contributions from  $\epem\to\qqbar$ (magenta dense-dot filled), $\TwoS\to X$ (green sparse-dot filled), $\OneS\to X$ (blue dotted), and $\OneS\to \gamma gg$ (red dashed) events.
}
\label{fig:bothMass}
\end{figure}

In total, $9.8\times10^3$ and $7.4\times10^6$ candidates satisfy the selection criteria in the low- and high-mass regions, respectively. The corresponding distributions of $m_X$ are shown in Fig.~\ref{fig:bothMass}. The backgrounds in the low-mass region consist of $\OneS\to\gamma gg$ (35\%); other \OneS decays, denoted $\OneS\to X$ (34\%); \TwoS decays without a dipion transition, denoted $\TwoS\to X$ (15\%); and $\epem\to\qqbar$ events (16\%). The corresponding background contributions in the high-mass region are 1\%, 66\%, 18\%, and 15\%. Background contributions from $\OneS\to\gamma gg$ decays reach a maximum near 5.5\gevcc and decrease above 7\gevcc.

We search for the $A^0$ resonance as a peak in the $m_X$ distribution. The reconstructed width of the $A^0$ is expected to strongly depend on its mass due to better photon energy resolution at lower photon energies. Therefore, an extended maximum likelihood fit in a local mass range is performed as a function of test-mass values, denoted $m_{A^0}$. For these fits, the parameters of the probability density function (PDF) used to model the shape of the signal distribution are fixed. The parameters of the background PDF, the number of signal events $N_{sig}$, and the number of background events are determined in the fit.

The signal $m_X$ PDF is modeled with a Crystal Ball function~\cite{ref:crys}, which consists of a Gaussian and a power-law component. The values of the signal PDF at a given value of $m_{A^0}$ are determined through interpolation from fits of simulated signal events at neighboring masses. The background $m_X$ PDF is modeled with a second-order polynomial.

\begin{figure}[h]
\includegraphics[width=\columnwidth]{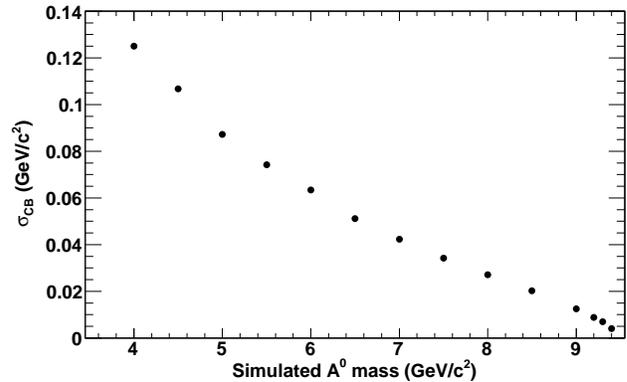}
\caption{
The $\sigma_{CB}$ parameter for $A^0$ decays of various simulated masses.
}
\label{fig:sigmaCB}
\end{figure}

The fits are performed to the $m_X$ spectrum, for various choices of $m_{A^0}$, in steps of 10 and 2~\mevcc for the low- and high-mass regions, respectively. The step sizes are at least 3 times smaller than the width of the signal $m_X$ PDF.  We use a local fitting range of $\pm$10$~\sigma_{CB}$ around $m_{A^0}$, where $\sigma_{CB}$ denotes the width of the Gaussian component of the Crystal Ball function. The $\sigma_{CB}$ parameter varies between 120 and 8~\mevcc for values of $m_{A^0}$ between 4.00 and 9.25\gevcc, as shown in Fig. \ref{fig:sigmaCB}. We do not perform a fit for $8.95 < m_{A^0} < 9.10$~\gevcc because of a large background from $\TwoS\to\gamma \chi_{bJ}(1P)$, $\chi_{bJ}(1P)\to \gamma \OneS$ decays.

The fitting procedure is validated using background-only pseudo-experiments.  The $m_X$ PDF used to generate pseudo-experiments for the low-mass region is obtained from a fit of a fifth-order polynomial to the low-mass region data.  The $m_X$ PDF used for the high-mass region is obtained from a fit of the sum of four exponential functions plus six Crystal Ball functions to the high-mass region data, with shape parameters fixed according to expectations from simulation and with the relative normalizations determined in the fit.  The Crystal Ball functions describe the $\TwoS\to\gamma \chi_{bJ}(1P)$ and $\chi_{bJ}(1P)\to \gamma \OneS$ transitions while the exponential terms describe the non-resonant background. Four exponential terms are used because the non-resonant background increases rapidly for higher $m_X$. The background fits are overlaid in Fig.~\ref{fig:bothMass}. The fitting procedure returns a null signal for most $m_{A^0}$ values but is found to require a correction to $N_{sig}$ for values of $m_{A^0}$ near 4.00 or 9.25~\gevcc. The corrections are determined from the average number of signal events found in the fits to the background-only pseudo-experiments. The corrections are applied as a function of $m_{A^0}$ and reach a maximum of $15$ and $50$ candidates in the low- and high-mass regions, respectively. The uncertainty of the correction is assumed to be half its value. 

The reconstruction efficiency takes into account the hadronization of the \ccbar system into $D$ mesons, the branching fraction of $D$ mesons to the five decay channels, detector acceptance, and the BDT selection. The efficiencies range from 4.0\% to 2.6\% for simulated $A^0$ masses between 4.00 and 9.25~\gevcc.

Potential bias introduced by the fitting procedure is evaluated using pseudo-experiments with different values of the product branching fraction $\mathcal{B}(\OneS\to \gamma A^0) \times \mathcal{B}(A^0 \to \ccbar$). For various choices of $m_{A^0}$, the extracted product branching fraction is found to be $(4\pm3)$\% higher than the value used to generate the events. This result is used to define a
  correction and its uncertainty.

Table \ref{tab:sysSummary} summarizes all correction factors and associated systematic uncertainties. The fit correction systematic uncertainty is added in quadrature with the statistical uncertainty of $N_{sig}$. All other correction factors are added in quadrature and applied to the reconstruction efficiency. A correction of 1.00 means we do not apply any correction but propagate the multiplicative uncertainty.

\begin{table}[h]
  	\begin{center}
 	\caption{Summary of corrections and their associated systematic uncertainties. All corrections are multiplicative except for the fit correction.}
	\begin{tabular}{c c c}
	\hline
	\hline
  	Source & Low region & High region \\
	\hline  	
	\hline
  	Fit correction (candidates) & up to $15\pm8$ & up to $50\pm25$ \\
  	BDT output modeling & $0.93\pm0.04$ & $0.98\pm0.01$ \\
	\hline  	
	\hline  	
	Source & \multicolumn{2}{c}{Both regions} \\
	\hline  	
	\hline  	
	$\ccbar$ hadronization & \multicolumn{2}{c}{$1.00\pm0.09$} \\
	Fit bias & \multicolumn{2}{c}{$1.04\pm0.03$} \\
	Dipion branching fraction & \multicolumn{2}{c}{$1.00\pm0.02$} \\
	Photon reconstruction & \multicolumn{2}{c}{$0.967\pm0.017$} \\
	$D$ mass resolution & \multicolumn{2}{c}{$0.98\pm0.01$} \\
	Finite simulation statistics & \multicolumn{2}{c}{$1.00\pm0.01$} \\
	\Y2S counting & \multicolumn{2}{c}{$1.00\pm0.01$} \\
	Dipion likelihood & \multicolumn{2}{c}{$1.02\pm0.01$} \\
	Dipion recoil mass & \multicolumn{2}{c}{$0.991\pm0.005$} \\
	\hline
	\hline
	\end{tabular}
      \label{tab:sysSummary}
        \end{center}
\end{table} 

The systematic uncertainties associated with the reconstruction efficiencies are dominated by the differences between data and simulation, including the BDT output modeling, \ccbar hadronization, $D$-candidate mass resolution, dipion recoil mass and likelihood modeling, and photon reconstruction. Other systematic uncertainties are associated with the fit bias (discussed above), the dipion branching fraction \cite{ref:pdg}, the finite size of the simulated signal sample, and the \TwoS counting \cite{Lees:2013rw}.

\begin{figure}[h]
\includegraphics[width=\columnwidth]{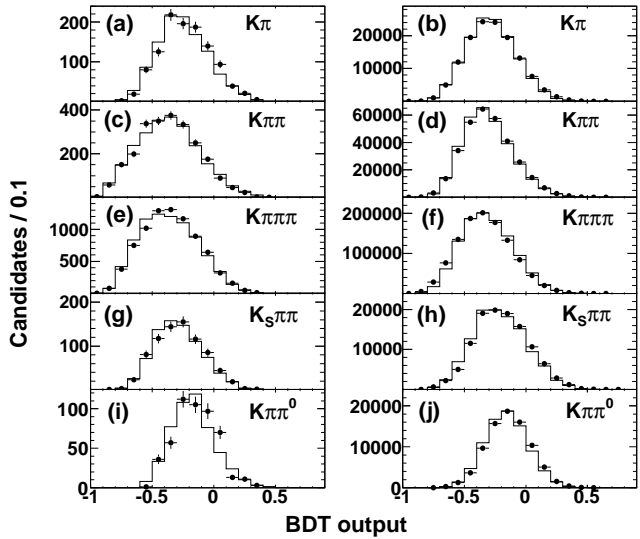}
\caption{
The BDT distributions in off-resonance data (points) and simulated $\epem\to\qqbar$ events (histograms) for the five $D$ meson decay modes. The results on the left (a, c, e, g, i) and right (b, d, f, h, j)  correspond to the low- and high-mass regions, respectively.
}
\label{fig:AllBDToff}
\end{figure}

The BDT output distributions in off-resonance data and $\epem\to\qqbar$ simulation, shown in Fig. \ref{fig:AllBDToff}, have consistent shapes but are slightly shifted from one another. The associated systematic uncertainty is estimated by shifting the simulated distributions so that the mean values agree with the data, and then recalculating the efficiencies. The reconstruction efficiencies decrease by 7\% and 2\% in the low- and high-mass regions, respectively.

The uncertainty associated with \ccbar hadronization is evaluated by comparing $D$ meson production in off-resonance data and $\epem\to\ccbar$ simulation normalized to the same luminosity. The difference in the yield varies from 1\% to 9\% for the five $D$ decay channels. We conservatively assign a global multiplicative uncertainty of 9\% that includes effects due to the hadronization modeling, particle identification, tracking, \piz reconstruction, and luminosity determination of the off-resonance data.

The uncertainty due to the discrepancy between the reconstructed $D$ mass resolution in data and simulation is estimated by Gaussian smearing of the $D$ mass input in simulation to match the data and measuring the difference in the reconstruction efficiency.

Further corrections to account for data and simulation differences in reconstruction efficiencies are estimated with similar methods. Corrections are applied to account for the dipion recoil mass reconstruction, the dipion likelihood modeling, and the photon reconstruction~\cite{ref:gammagamma}.

\begin{figure}[h]
\includegraphics[width=\columnwidth]{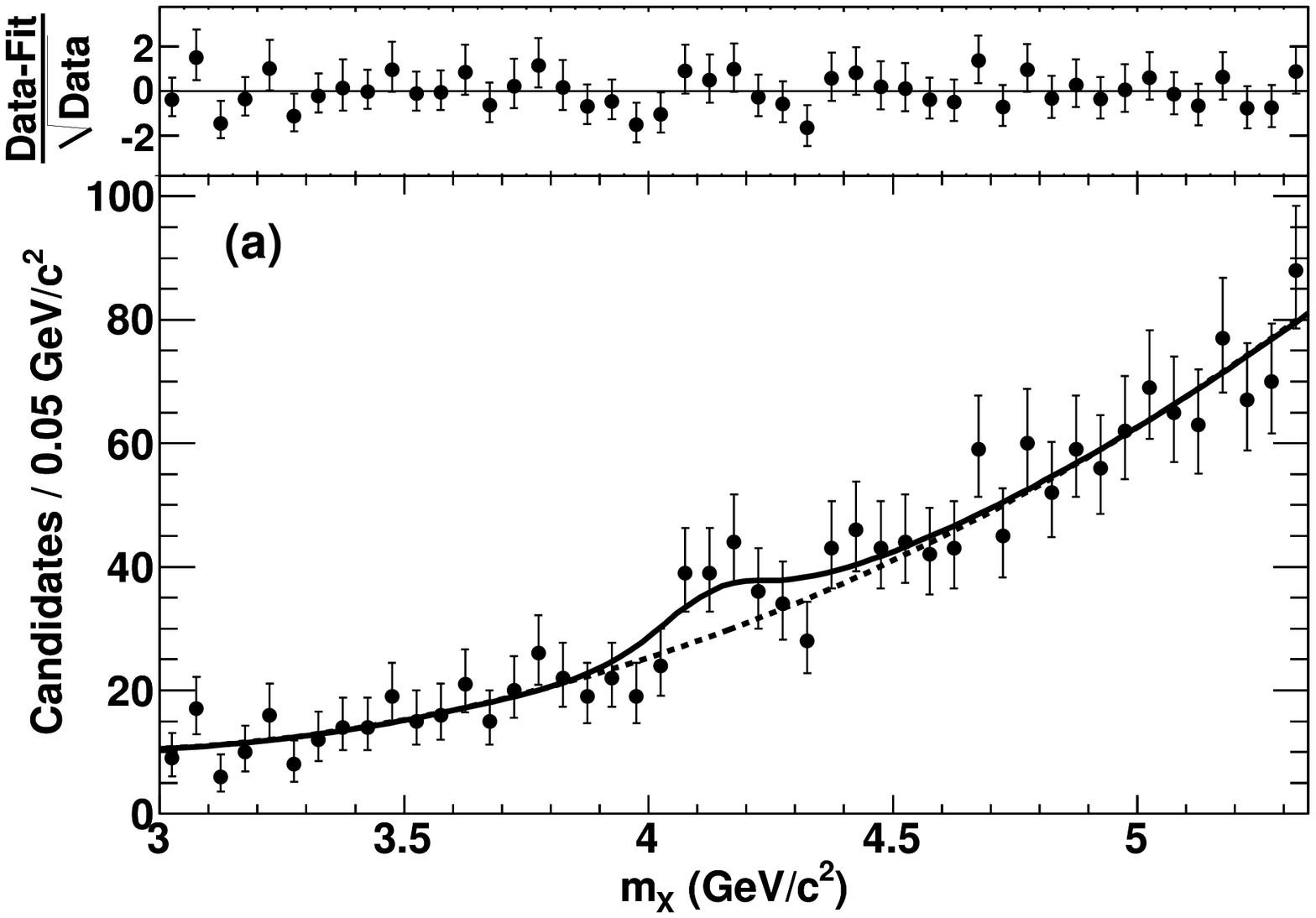}
\includegraphics[width=\columnwidth]{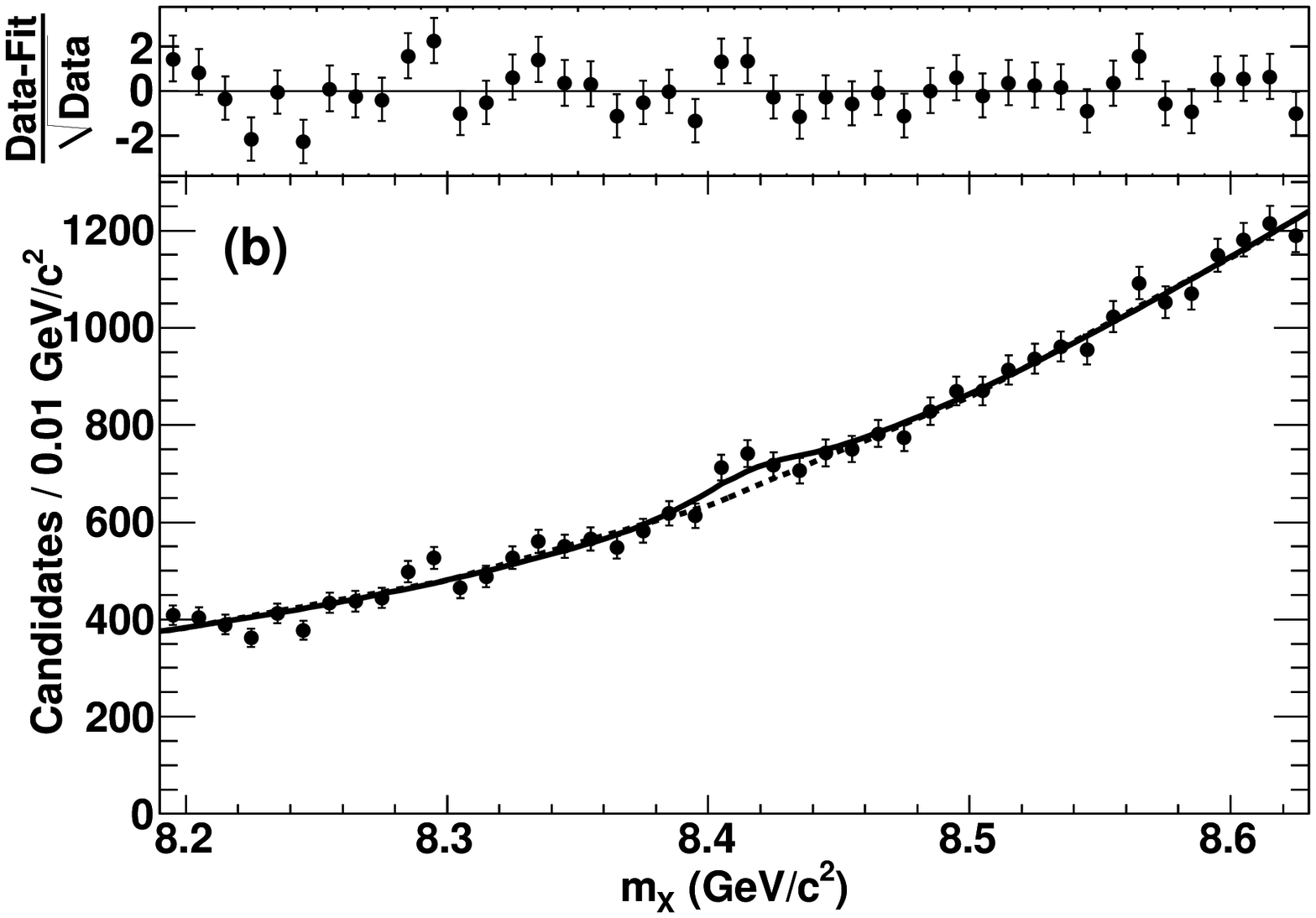}
\caption{
The fits with the highest local significance in the low- (a) and high- (b) mass regions. The solid line is the fit that includes a signal. The dotted line is the background component of the solid line.
}
\label{fig:bothHighest}
\end{figure}

The highest observed local significance in the low-mass region is 2.3 standard deviations, including statistical uncertainties only, at 4.145~\gevcc. The corresponding result for the high-mass region is 2.0 standard deviations at 8.411~\gevcc. The fits are shown in Fig.~\ref{fig:bothHighest}. Such fluctuations occur in 54\% and 80\% of pseudo-experiments, respectively. Hence our data are consistent with the background-only hypothesis.

\begin{figure}[h]
\includegraphics[width=\columnwidth]{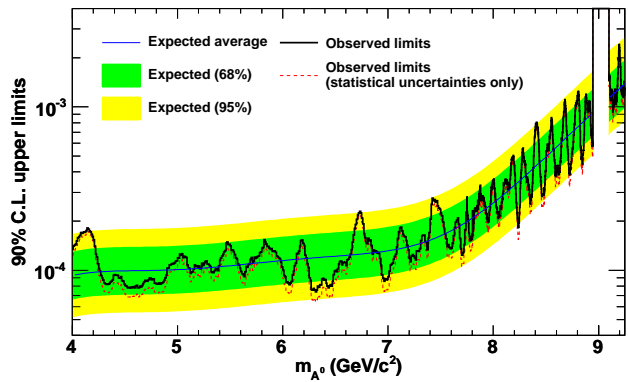}
\caption{
(color online) The 90\% C.L. upper limits on the product branching fraction $\mathcal{B}(\OneS\to \gamma A^0) \times \mathcal{B}(A^0 \to \ccbar$) using all uncertainties (thick line) and using statistical uncertainties only (thin dashed line). The inner and outer bands contain 68\% and 95\% of our expected upper limits. The bands are calculated using all uncertainties. The thin solid line in the center of the inner band is the expected upper limit.
}
\label{fig:BFUL}
\end{figure}

Upper limits on the product branching fraction $\mathcal{B}(\OneS\to \gamma A^0) \times \mathcal{B}(A^0 \to \ccbar$) at 90\% confidence level (C.L.) are determined assuming a uniform prior, with the constraint that the product branching fraction be greater than zero. The distribution of the likelihood function for $N_{sig}$ is assumed to be Gaussian with a width equal to the total uncertainty in $N_{sig}$. The upper limits obtained from the low-mass region are combined with those from the high-mass region to define a continuous spectrum for the upper limits.  The results are shown in Fig.~\ref{fig:BFUL}.

In summary, we search for a resonance in radiative decays of the \OneS with a charm tag. We do not observe a significant signal and set upper limits on the product branching fraction $\mathcal{B}(\OneS\to \gamma A^0) \times \mathcal{B}(A^0 \to \ccbar$) ranging from $7.4 \times 10^{-5}$ to $2.4 \times 10^{-3}$ for $A^0$ masses from 4.00 to 9.25 \gevcc, excluding masses from 8.95 to 9.10 \gevcc because of background from $\TwoS\to\gamma \chi_{bJ}(1P)$, $\chi_{bJ}(1P)\to \gamma \OneS$ decays. These results will further constrain the NMSSM parameter space.

We are grateful for the 
extraordinary contributions of our \pep2\ colleagues in
achieving the excellent luminosity and machine conditions
that have made this work possible.
The success of this project also relies critically on the 
expertise and dedication of the computing organizations that 
support \babar.
The collaborating institutions wish to thank 
SLAC for its support and the kind hospitality extended to them. 
This work is supported by the
US Department of Energy
and National Science Foundation, the
Natural Sciences and Engineering Research Council (Canada),
the Commissariat \`a l'Energie Atomique and
Institut National de Physique Nucl\'eaire et de Physique des Particules
(France), the
Bundesministerium f\"ur Bildung und Forschung and
Deutsche Forschungsgemeinschaft
(Germany), the
Istituto Nazionale di Fisica Nucleare (Italy),
the Foundation for Fundamental Research on Matter (The Netherlands),
the Research Council of Norway, the
Ministry of Education and Science of the Russian Federation, 
Ministerio de Econom\'{\i}a y Competitividad (Spain), the
Science and Technology Facilities Council (United Kingdom),
and the Binational Science Foundation (U.S.-Israel).
Individuals have received support from 
the Marie-Curie IEF program (European Union) and the A. P. Sloan Foundation (USA). 

% NOTES:
% add "and the Binational Science Foundation (U.S.-Israel)"  07-Oct-2013 Bill Gary (Abi Soffer request)

\end{document}